\documentclass[prd,aps,nofootinbib,superscriptaddress]{revtex4}
\usepackage{epsfig}
\usepackage{amsmath}
\usepackage{amsfonts}
\usepackage{amssymb}
\usepackage{graphicx}
\usepackage{graphics}
\usepackage{hyperref}
\usepackage{epstopdf}
\begin{document}
\title{A Three-Loop Model of Neutrino Mass with Dark Matter}
\author{Amine Ahriche}
\email{aahriche@ictp.it}
\affiliation{Department of Physics, University of Jijel, PB 98 Ouled Aissa, DZ-18000 Jijel, Algeria.}
\affiliation{The Abdus Salam International Centre for Theoretical Physics, Strada Costiera
11, I-34014, Trieste, Italy.}
\author{Chian-Shu~Chen}
\email{chianshu@phys.sinica.edu.tw}
\affiliation{Physics Division, National Center for Theoretical Sciences, Hsinchu, 300 Taiwan.}
\affiliation{Department of Physics, National Tsing Hua University, Hinschu, 300 Taiwan.}
\author{Kristian~L. McDonald}
\email{klmcd@physics.usyd.edu.au} \affiliation{ARC Centre of
Excellence for Particle Physics at the Terascale, School of Physics,
The University of Sydney, NSW 2006, Australia.}
\author{Salah Nasri}
\email{snasri@uaeu.ac.ae}
\affiliation{Physics Department, UAE University, POB 17551, Al Ain, United Arab Emirates.}
\affiliation{Laboratoire de Physique Theorique, ES-SENIA University, DZ-31000 Oran, Algeria.}

\begin{abstract}
We propose a model in which the origin of neutrino mass is dependent on the
existence of dark matter. Neutrinos acquire mass at the three-loop level and
the dark matter is the neutral component of a fermion triplet. We show that
experimental constraints are satisfied and that the dark matter can be tested
in future direct-detection experiments. Furthermore, the model predicts a
charged scalar that can be within reach of collider experiments like the LHC.

\end{abstract}
\maketitle

\section{Introduction\label{sec:introduction}}

The observed neutrino mixing provides concrete evidence that the Standard
Model (SM) is incomplete. Similarly, the need to explain dark matter (DM)
motivates the addition of a long-lived particle species to the SM. It is
natural to ask if these two short-comings of the SM could have a unified
explanation; are the DM and neutrino mass problems related?

A simple model connecting the origin of neutrino mass to the existence of DM
was proposed by Krauss, Nasri and Trodden (KNT)~\cite{Krauss:2002px}. The
basic idea was to extend the SM to include new fields, one of which was the
DM, such that neutrino mass was radiatively generated at the three-loop level
(for detailed studies see
Refs.~\cite{Baltz:2002we,Cheung:2004xm,Ahriche:2013zwa,Ahriche:2014xra}). In
this model DM played a key role in enabling neutrino mass; if the DM is
removed the loop diagram simply does not manifest and neutrinos remain massless

In recent years, a number of alternative models were proposed that
similarly predicted relationships between radiative neutrino mass
and DM (for a review see Ref. \cite{SB}). A particularly simple
one-loop model was proposed by Ma~\cite{Ma:2006km} and further
studied in Ref.~\cite{Ho:2013hia}. Other related models also
appeared~\cite{Aoki:2008av,Kajiyama:2013zla,Kanemura:2011mw,Ahriche:2014oda,MarchRussell:2009aq},
including a colored version of the KNT model which employed
leptoquarks~\cite{Ng:2013xja}. In both the KNT and Ma models the SM
is extended to include gauge-singlet fermions. Normally these
fermions couple to SM leptons and generate neutrino mass at
tree-level. However, in the models of KNT and Ma the coupling to the
SM does not occur, due to a discrete symmetry. This avoids a
tree-level (Type-I) seesaw mechanism and ensures DM longevity.

It is well known that the seesaw mechanism can be generalized to a triplet
(Type-III) variant by using $SU(2)_{L}$-triplet fermions, instead of singlet
fermions~\cite{Foot:1988aq}. Similarly, the Ma model can be generalized to a
triplet variant~\cite{Ma:2008cu}. In this paper we present a new model
motivating a connection between the origin of neutrino mass and DM. The model
is essentially a triplet variant of the KNT model, and employs triplet
fermions to generate radiative masses at the three-loop level. We show that
viable neutrino masses are obtained, and that constraints from flavor-changing
decays, the anomalous magnetic moment of the muon, and neutrino-less
double-beta decay can be satisfied. Viable DM is also obtained, in the form of
the neutral component of a triplet fermion. This candidate should produce
signals in next-generation direct-detection experiments. The model contains a
new charged scalar that can also be within reach of collider experiments.

The plan of this paper is as follows. In Section~\ref{sec:nu_mass} we describe
the model and detail the origin of neutrino mass. Various constraints are
analyzed in Section~\ref{sec:constraints} and we consider DM in
Section~\ref{sec:dark_matter}. Conclusions appear in Section~\ref{sec:conc}.

\section{Three-Loop Radiative Neutrino Masses\label{sec:nu_mass}}

\subsection{The Model}

The SM contains the gauge symmetry $SU(3)_{c}\times SU(2)_{L}\times U(1)_{Y}$
. In this work we extend the particle spectrum of the SM to include a charged
scalar, $S^{+}\sim(1,1,2)$, a triplet scalar, $T\sim(1,3,2)$, and real fermion
triplets, $E_{i}\sim(1,3,0)$, where $i=1,2,3$ labels generations. A $Z_{2}$
symmetry with action $\{T,E_{i}\}\rightarrow\{-T,-E_{i}\}$ is also imposed,
while all other fields are $Z_{2}$ even. The $SU(2)_{L}$ triplet fields are
written as symmetric matrices, $T_{ab}$ and $E_{ab}$, with components
\begin{align}
& T_{11}=T^{++},\quad T_{12}=T_{21}=\frac{1}{\sqrt{2}}\,T^{+},\quad
T_{22}=T^{0},\nonumber\\
& \ E_{11}=E_{L}^{+},\quad E_{12}=E_{21}=\frac{1}{\sqrt{2}}\,E_{L}^{0},\quad
E_{22}=E_{L}^{-}\equiv(E_{R}^{+})^{c}.
\end{align}
With these elements, the Lagrangian includes the terms
\begin{equation}
\mathcal{L}\supset\mathcal{L}_{\text{{\tiny SM}}}+\{f_{\alpha\beta}%
\,\overline{L_{\alpha}^{c}}\,L_{\beta}\,S^{+}+g_{i\alpha}\,\overline{E_{i}%
}\,T\,e_{\alpha R}+\mathrm{H.c}\}\;-\,\frac{1}{2}\,\overline{E_{i}^{c}%
}\,\mathcal{M}_{ij}\,E_{j}\;-\;V(H,S,T),\label{eq:lagragngian}%
\end{equation}
where $L_{\alpha}\sim(1,2,-1)$ and $e_{\alpha R}\sim(1,1,-2)$ are the SM
leptons and $f_{\alpha\beta}=-f_{\beta\alpha}$ are Yukawa couplings. Lowercase
greek letters label lepton flavors, $\alpha,\,\beta\in\{e,\,\mu,\,\tau\}$. The
singlet leptons couple to the exotics through a Yukawa matrix $g_{i\alpha}$,
and the superscript \textquotedblleft$c$" denotes charge conjugation.

Due to the discrete symmetry the triplet fermions do not mix with the SM at
any order in perturbation theory. The triplet mass term gives
\begin{equation}
-(\overline{E_{i}^{c}})_{ab}\,\mathcal{M}_{ij}\,(E_{j})_{cd}\,\epsilon
^{ac}\,\epsilon^{bd}=-\{\overline{E_{iR}^{+}}\,\mathcal{M}_{ij}\,E_{jL}%
^{+}-\frac{1}{2}\overline{(E_{iL}^{0})^{c}}\,\mathcal{M}_{ij}\,E_{jL}%
^{0}\}.\label{eq:mass_lagrangeE}%
\end{equation}
Without loss of generality we work in a diagonal basis with $\mathcal{M}%
_{ij}=\mathrm{diag}(M_{1},M_{2},M_{3})$, where $M_{1}$ is the lightest
triplet-fermion mass. The charged and neutral components of $E_{i}$ are
degenerate at tree-level, though radiative corrections involving SM gauge
bosons lift this degeneracy, making the charged component heavier. For
$M_{E}\sim$~TeV the mass splitting is $\Delta M_{E}\simeq167$%
~MeV~\cite{Cirelli:2005uq}. For the most part we can neglect this small
splitting. To bring the neutral fermion mass-term to the correct sign, one
defines the Majorana fermions as $E_{i}^{0}=E_{i,L}^{0}-(E_{i,L}^{0})^{c}$
(see Appendix~\ref{app:chiral}). The DM candidate is the lightest neutral
fermion $E_{1}^{0}$ with mass $M_{DM}=M_{1}$. When the mass splitting is
neglected, one can denote the mass for all members of the lightest triplet
simply as $M_{DM}$.

\subsection{Neutrino Mass}

\begin{figure}[t]
\begin{center}
\includegraphics[width = 0.4\textwidth]{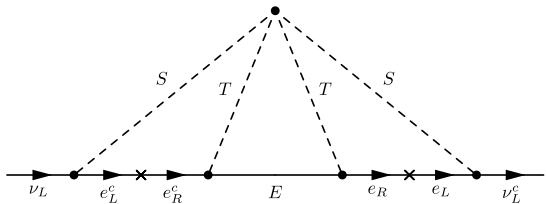}
\end{center}
\caption{Three-loop diagram for radiative neutrino mass. Here $S\sim(1,1,2)$
and $T\sim(1,3,2)$ are scalars while $E\sim(1,3,0)$ is a fermionic triplet.
There are three distinct diagrams with the sets $\{T^{+},E^{0},T^{-}\}$,
$\{T^{++},(E^{+})^{c},T^{0}\}$ and $\{T^{0},E^{+},T^{--}\}$ propagating in the
inner loop.}%
\label{fig:3loop_nuDM}%
\end{figure}

The scalar potential contains the terms
\begin{equation}
V(H,S,T)\supset\frac{\lambda_{S}}{4}(S^{-})^{2}T_{ab}T_{cd}\epsilon
^{ac}\epsilon^{bd}+\frac{\lambda_{S}^{\ast}}{4}(S^{+})^{2}(T^{\ast}%
)^{ab}(T^{\ast})^{cd}\epsilon_{ac}\epsilon_{bd},
\end{equation}
which, in combination with the new Yukawa Lagrangian shown in
Eq.~\eqref{eq:lagragngian}, explicitly break lepton number symmetry.
Therefore, the three vertices should appear simultaneously in the Majorana
mass diagram. As a result, neutrino masses are generated at the three-loop
level, as shown in Figure~\ref{fig:3loop_nuDM}, where there are three distinct diagrams, corresponding to the sets
$\{T^{+},E^{0},T^{-}\}$, $\{T^{++},(E^{+})^{c},T^{0}\}$ and $\{T^{0}%
,E^{+},T^{--}\}$ propagating in the inner loop. Note that the intermediate
fermion is a triplet instead of the singlet field in the KNT model. The
triplet scalars are nondegenerate at tree-level due to the term $\lambda
_{HT}H^{\dagger}TT^{\dagger}H\subset V(H,S,T)$. However, this splitting is
small for the parameter space of interest here, provided $|\lambda_{HT}|<1$.
Thus, one can neglect the mass-splitting among members of both the scalar and
fermion triplets when calculating the mass diagram. The calculation gives
\begin{equation}
(\mathcal{M}_{\nu})_{\alpha\beta}=\frac{3\lambda_{S}}{(4\pi^{2})^{3}}%
\frac{m_{\gamma}m_{\delta}}{M_{T}}\,f_{\alpha\gamma}\,f_{\beta\delta
}\,g_{\gamma i}^{\ast}\,g_{\delta i}^{\ast}\times F\left( \frac{M_{i}^{2}%
}{M_{T}^{2}},\frac{M_{S}^{2}}{M_{T}^{2}}\right) .\label{mth}%
\end{equation}
Here $m_{\gamma,\delta}$ denote SM charged-lepton masses, and $F(x,y)$ is a
function encoding the loop integrals (see Appendix \ref{app:loop_function}).
$M_{S}$ is the charged-singlet mass and $M_{\text{{\tiny T}}}$ is the triplet
scalar mass.

The elements of the neutrino mass matrix can be related to the mass
eigenvalues and the Pontecorvo-Maki-Nakawaga-Sakata (PMNS) mixing
matrix~\cite{Pontecorvo:1967fh} elements in the standard way
\begin{equation}
(\mathcal{M}_{\nu})_{\alpha\beta}=[U_{\nu}\cdot\mathrm{diag}(m_{1}%
,\,m_{2},\,m_{3})\cdot U_{\nu}^{\dag}]_{\alpha\beta}.\label{mexp}%
\end{equation}
The PMNS matrix is parametrized as
\begin{equation}
U_{\nu}=\left(
\begin{array}
[c]{ccc}%
c_{12}c_{13} & c_{13}s_{12} & s_{13}e^{-i\delta_{D}}\\
-c_{23}s_{12}-c_{12}s_{13}s_{23}e^{i\delta_{D}} & c_{12}c_{23}-s_{12}%
s_{13}s_{23}e^{i\delta_{D}} & c_{13}s_{23}\\
s_{12}s_{23}-c_{12}c_{23}s_{13}e^{i\delta_{D}} & -c_{12}s_{23}-c_{23}%
s_{12}s_{13}e^{i\delta_{D}} & c_{13}c_{23}%
\end{array}
\right) \times U_{p},
\end{equation}
where $U_{p}=\mathrm{diag}(1,\,e^{i\theta_{\alpha}/2},\,e^{i\theta_{\beta}%
/2})$ contains the Majorana phases $\theta_{\alpha}$ and $\theta_{\beta}$,
$\delta_{D}$ is the Dirac phase, and $s_{ij}\equiv\sin\theta_{ij}$,
$c_{ij}\equiv\cos\theta_{ij}$. The best-fit experimental values for the mixing
angles and mass-squared differences are $s_{12}^{2}=0.320_{-0.017}^{+0.016}$,
$s_{23}^{2}=0.43_{-0.03}^{+0.03}$, $s_{13}^{2}=0.025_{-0.003}^{+0.003}$,
$\Delta m_{21}^{2}=7.62_{-0.19}^{+0.19}\times10^{-5}\mathrm{eV}^{2}$, and
$|\Delta m_{13}^{2}|=2.55_{-0.09}^{+0.06}\times10^{-3}\mathrm{eV}^{2}%
$~\cite{Tortola:2012te}. Within these ranges, one can determine the regions of
parameters space where viable neutrino masses are compatible with experimental
constraints, and in agreement with the measured DM relic density.

\section{Experimental Constraints\label{sec:constraints}}

We shall discuss DM in detail in Section~\ref{sec:dark_matter}. For the moment
we note that the lightest $Z_{2}$-odd field is a stable DM candidate. There
are two possibilities for the DM, namely $T^{0}$ and $E_{1}^{0}$. However, $T$
has nonzero hypercharge so that $T^{0}$ couples to the $Z$ boson. As the
CP-even and CP-odd components of $T^{0}$ are mass-degenerate, stringent
direct-detection constraints apply and, in fact, exclude $T^{0}$ as a DM
candidate \cite{Araki:2011hm}. Consequently the neutral fermion $E_{1}^{0}$ is
the only viable DM candidate in the model and $M_{{\tiny T}}>M_{{\tiny DM}}$
is required to ensure DM stability. In earlier models with triplet-fermion DM
\cite{Ma:2008cu,Cirelli:2005uq}, the observed relic density was obtained via
$SU(2)_{L}$ (co-)annihilation channels and required a DM mass around $M_{{DM}%
}\sim2.3-2.4$~TeV. However, we shall see that in our model the additional
annihilation channels require the DM mass to be in the range $M_{{\tiny DM}%
}\sim2.35-2.75$~TeV. Thus, both $T$ and $E_{i}$ are too heavy to be produced
at the LHC, though the $Z_{2}$-even field $S$ could be within the reach of the
LHC, as we discuss below.

\begin{figure}[t]
\centering
\begin{tabular}
[c]{ccc}%
\epsfig{file=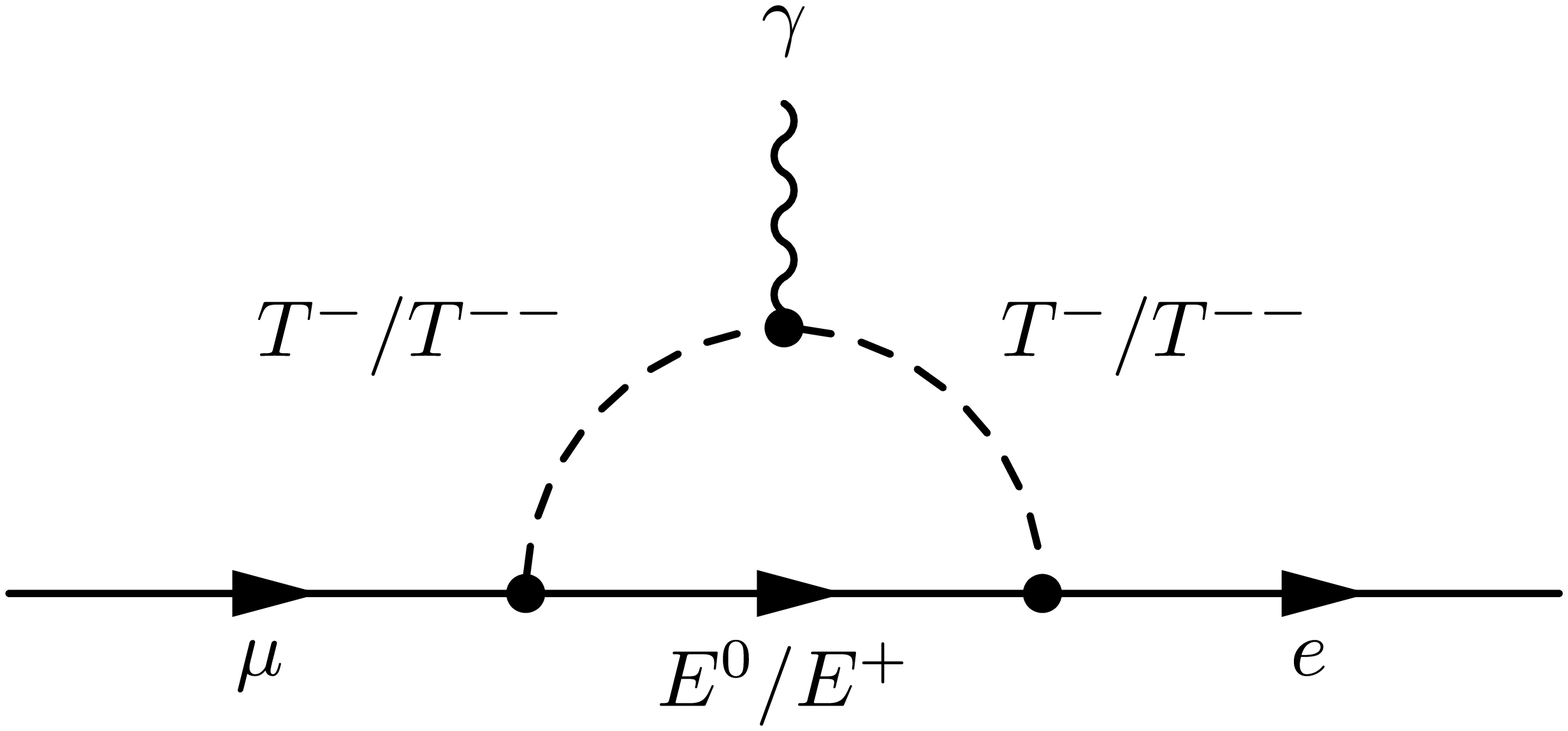,width=0.4\linewidth,clip=} & &
\epsfig{file=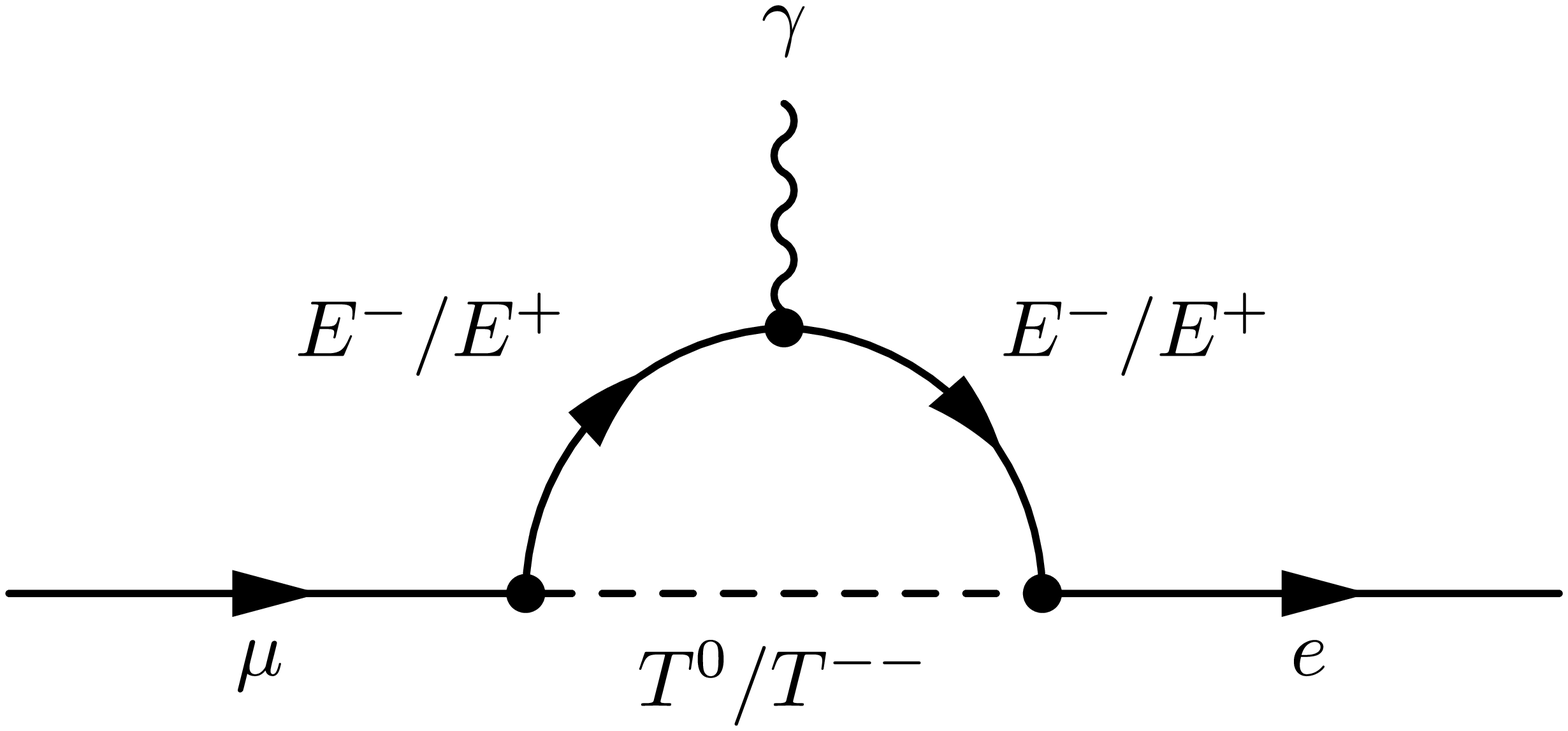,width=0.4\linewidth,clip=}\\
(a) & & (b)\\
& &
\end{tabular}
\caption{Diagrams for $\mu\rightarrow e+\gamma$ due to the $Z_{2}$-odd fields
$E\sim(1,3,0)$ and $T\sim(1,3,2)$. Two additional diagrams, with the photon
attached to the lower line, and a diagram due the singlet scalar
$S\sim(1,1,2)$ are also shown.}%
\label{fig:muEgamma}%
\end{figure}

The Yukawa couplings $g_{i\alpha}$ induce flavor changing processes like
$\mu\rightarrow e+\gamma$. At the one-loop level the exotic triplets give four
distinct diagrams, as shown in Figure~\ref{fig:muEgamma}. For two of these
diagrams the photon is attached to the internal fermion line. In the limit
that the mass-splitting between $T^{0}$ and $T^{--}$ vanishes, the amplitudes
for these diagrams differ by an overall sign. Thus, in this limit, the
coherent sum of these amplitudes vanishes, and to good approximation the
diagrams can be neglected. Therefore we need only calculate the diagrams in
Figure~\ref{fig:muEgamma}a. There is also a diagram mediated by the singlet
scalar $S$ that must be included. Putting all this together, the branching
ratio for $\mu\rightarrow e+\gamma$ is%
\begin{align}
B(\mu\rightarrow e\gamma) & =\frac{\Gamma(\mu\rightarrow e+\gamma)}%
{\Gamma(\mu\rightarrow e+\nu+\bar{\nu})}\nonumber\\
& \simeq\frac{\alpha\upsilon^{4}}{384\pi}\times\left\{
\frac{|f_{\mu\tau }f_{\tau
e}^{\ast}|^{2}}{M_{S}^{4}}+\frac{324}{M_{T}^{4}}\left\vert \sum
_{i}g_{ie}^{\ast}g_{i\mu}F_{2}(M_{i}^{2}/M_{T}^{2})\right\vert
^{2}\right\}
,\label{eq:muEgamma_BF}%
\end{align}
where $F_{2}(R)=[1-6R+3R^{2}+2R^{3}-6R^{2}\log R]/[6(1-R)^{4}]$.
Eq.~\eqref{eq:muEgamma_BF} can also be used to determine $B(\tau\rightarrow
\mu+\gamma)$ by simply changing flavor labels.

Replacing the final-state electron with a muon, the diagrams in
Figure~\ref{fig:muEgamma} contribute to the muon magnetic moment. Similar
arguments hold for the calculation of the muon magnetic moment; the diagrams
with the photon attached to the internal fermion cancel when the mass
splitting for the scalar triplet is neglected. The result for the muon
anomalous magnetic moment is
\begin{equation}
\delta a_{\mu}=\frac{m_{\mu}^{2}}{16\pi^{2}}\left\{ \sum_{\alpha\neq\mu}%
\frac{|f_{\mu\alpha}|^{2}}{6M_{S}^{2}}+\sum_{i}\frac{3|g_{i\mu}|^{2}}%
{M_{T}^{2}}F_{2}(M_{i}^{2}/M_{T}^{2})\right\} .\label{amu}%
\end{equation}

\begin{figure}[t]
\begin{center}
\includegraphics[width = 0.5\textwidth]{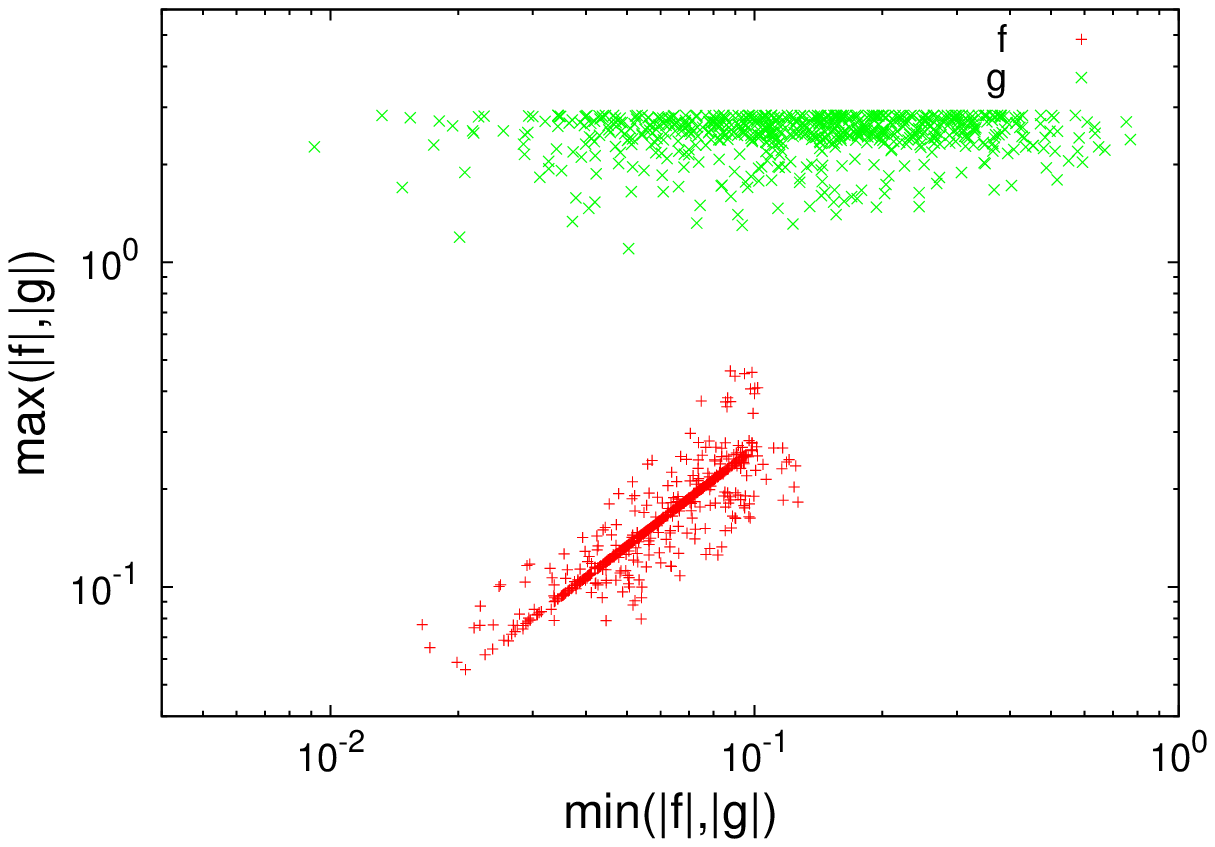}~\includegraphics[width
=0.5\textwidth]{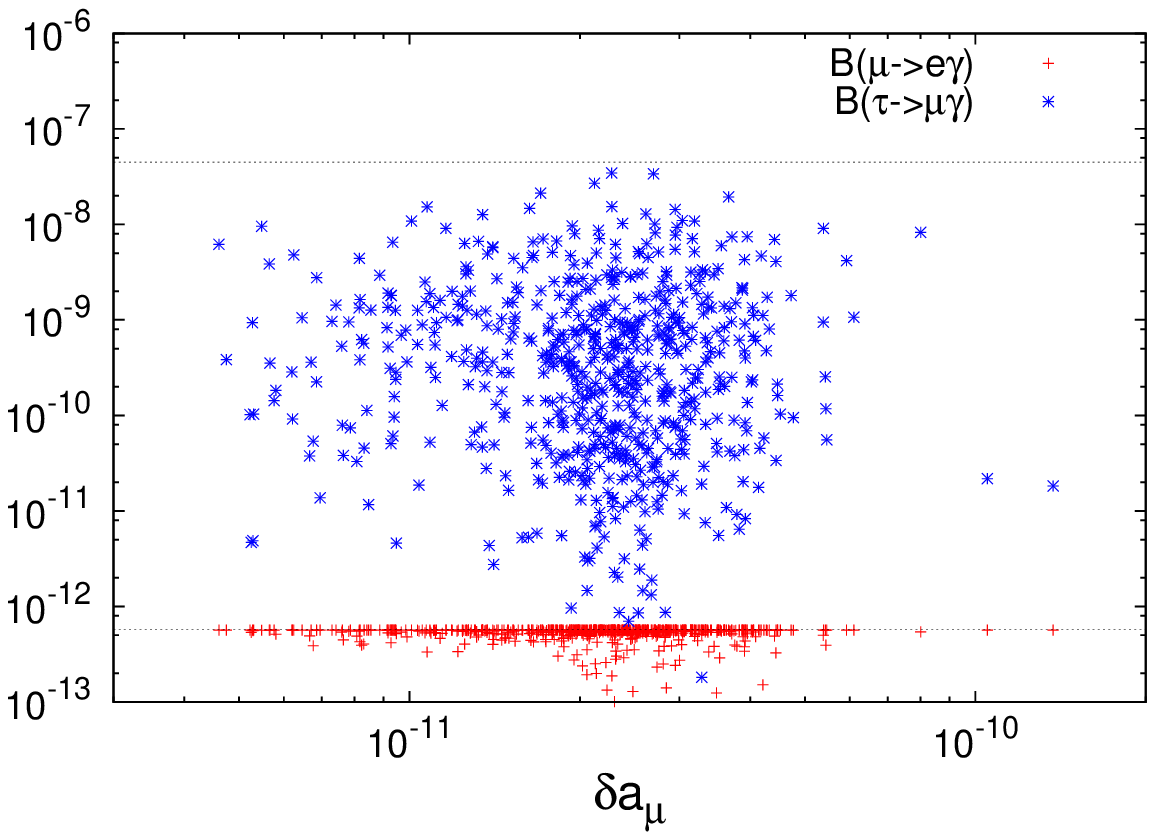}
\end{center}
\caption{Left: Viable regions of parameter space for the Yukawa couplings
$f_{\alpha\beta}$ and $g_{i\alpha}$. Here, neutrino mass/mixing and lepton
flavor-changing constraints are satisfied, and the correct DM relic abundance
is obtained. Right: Branching fractions for lepton flavor violating decays
$\mu\rightarrow e+\gamma$ and $\tau\rightarrow\mu+\gamma$ versus the anomalous
magnetic moment of the muon. Branching fraction limits appear as horizontal
lines and all points easily satisfy the magnetic moment constraint (which is
too large to appear in the figure).}%
\label{LFV}%
\end{figure}

After matching the calculated neutrino mass-matrix elements with the neutrino
mixing data, one finds a significant region of parameter space that is
consistent with low-energy constraints, as shown in Figure~\ref{LFV}.
We find that the Yukawa couplings $f_{\alpha\beta}$ lie in the
range $\left\vert f_{\alpha\beta}\right\vert \sim0.02-0.5$, and the couplings
$g_{i\alpha}$ are expected to be $\mathcal{O}(1)$. We restrict the latter to
the perturbative region with $|g_{i\alpha}|\lesssim3$. From the left panel,
one notices that the constraint from $\mu\rightarrow e+\gamma$ is most severe.
The null-results from searches for neutrino-less double-beta decay provide an
additional constraint of $(\mathcal{M}_{\nu})_{ee}\lesssim0.35$%
~eV~\cite{Simkovic:2009pp}, though this is easily satisfied. Next generation
experiments will improve this bound to the level of $(\mathcal{M}_{\nu}%
)_{ee}\lesssim0.01$~eV~\cite{Avignone:2005cs,Rodejohann:2011mu}.

We note that when considering only one or two generations of the triplet
fermions, no solution that simultaneously accommodates the neutrino mass and
mixing data, low-energy flavor physics constraints, and the DM relic density,
could be found. Thus a minimum of three generations of exotic triplet-fermions
are required.

\section{Dark Matter\label{sec:dark_matter}}

\subsection{Relic Density}

The lightest neutral fermion, $E_{1}^{0}$, is a stable DM candidate. There are
two classes of interactions that can maintain thermal contact between the DM
and the SM in the early Universe. The first class of interactions are mediated
by the triplet scalar, while the second involve $SU(2)_{L}$ gauge
interactions. In addition to annihilation processes like $E^{0}E^{0}%
\rightarrow SM$, there are coannihilations like $E^{0}E^{\pm}\rightarrow SM$.
Due to the small mass-splitting between neutral and charged triplet-fermions,
and the fact that coannihilation cross sections can be larger than
annihilation ones, coannihilations cannot be neglected.

The annihilation channels mediated by triplet scalars give
\begin{equation}
\sigma(2E^{0}\rightarrow\ell_{\beta}^{+}\ell_{\alpha}^{-})\times v_{r}%
=\frac{|g_{1\alpha}^{\ast}g_{1\beta}|^{2}}{48\pi}\frac{M_{DM}^{2}(M_{DM}%
^{4}+M_{T}^{4})}{(M_{DM}^{2}+M_{T}^{2})^{4}}\times v_{r}^{2}\ \equiv
\ \sigma_{00}^{\alpha\beta}\times v_{r},
\end{equation}
where $v_{r}$ is the relative velocity of DM particles in the center-of-mass
frame. As expected for Majorana DM, there are no $s$-wave annihilations in the
limit where final-state fermion masses are neglected. There are no
coannihilations mediated by $T$ but one must include the charged fermion
annihilations:
\begin{equation}
\sigma(E^{-}E^{+}\rightarrow\ell_{\beta}^{+}\ell_{\alpha}^{-})\times
v_{r}=\frac{|g_{1\alpha}^{\ast}g_{1\beta}|^{2}}{48\pi}\frac{M_{DM}^{2}%
(M_{DM}^{4}+M_{T}^{4})}{(M_{DM}^{2}+M_{T}^{2})^{4}}\times v_{r}^{2}%
\ \equiv\ \sigma_{-+}^{\alpha\beta}\times v_{r}.
\end{equation}
For the $SU(2)_{L}$ channels we work in the $SU(2)_{L}$-symmetric limit,
neglecting gauge-boson masses. This is justified \emph{a~posteriori} as we
find $M_{{DM}}\sim$~TeV is required. We can therefore use standard results in
the literature~\cite{Cirelli:2009uv}.

Ignoring the tiny mass-splitting, we combine annihilation and coannihilation
channels in the standard way~\cite{Griest:1990kh}, giving
\begin{equation}
\sigma_{eff}(2E\rightarrow SM)\times
v_{r}=\frac{1}{g_{eff}^{2}}\left[ \sigma_{W}\times
v_{r}+\sum_{\alpha,\beta}\left\{ g_{0}^{2}\,\sigma
_{00}^{\alpha\beta}+2g_{\pm}\,\sigma_{-+}^{\alpha\beta}\right\}
\times v_{r}\right] .
\end{equation}
Here $g_{0}=g_{\pm}=2$ and $g_{eff}=g_{0}+2g_{\pm}$. The $SU(2)$ channels are
denoted by
\begin{equation}
\sigma_{W}\equiv\sigma(2E\rightarrow_{W}SM)\ \simeq\ \frac{\pi\alpha_{2}^{2}%
}{2M_{DM}^{2}v_{r}}\left\{ 222+\frac{51}{2}v_{r}^{2}\right\} ,
\end{equation}
while the $T$-exchange cross sections are defined above.\footnote{If $p$-wave
processes are neglected and $T$-exchange channels ignored, our result matches
the $s$-wave expression of Ref.~\cite{Ma:2008cu}.} \begin{figure}[t]
\begin{center}
\includegraphics[width = 0.5\textwidth]{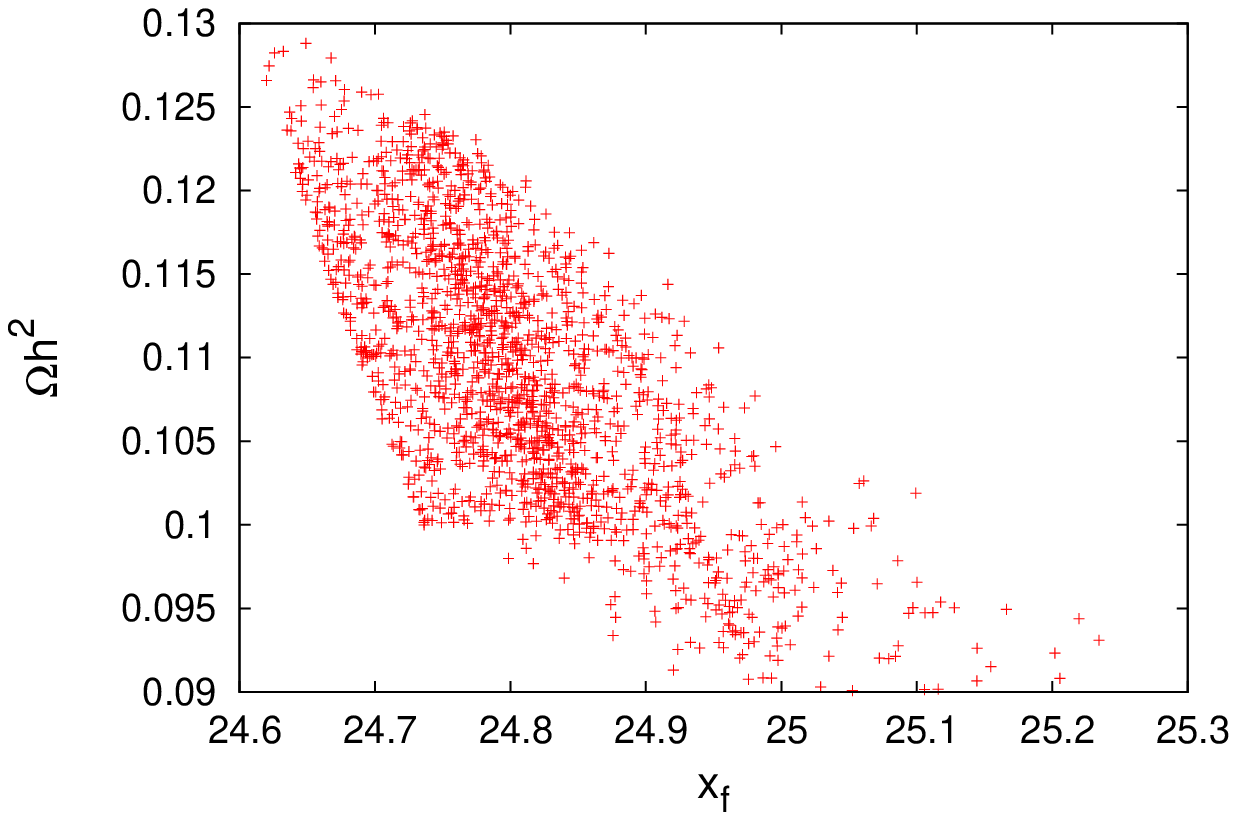}~\includegraphics[width =
0.5\textwidth]{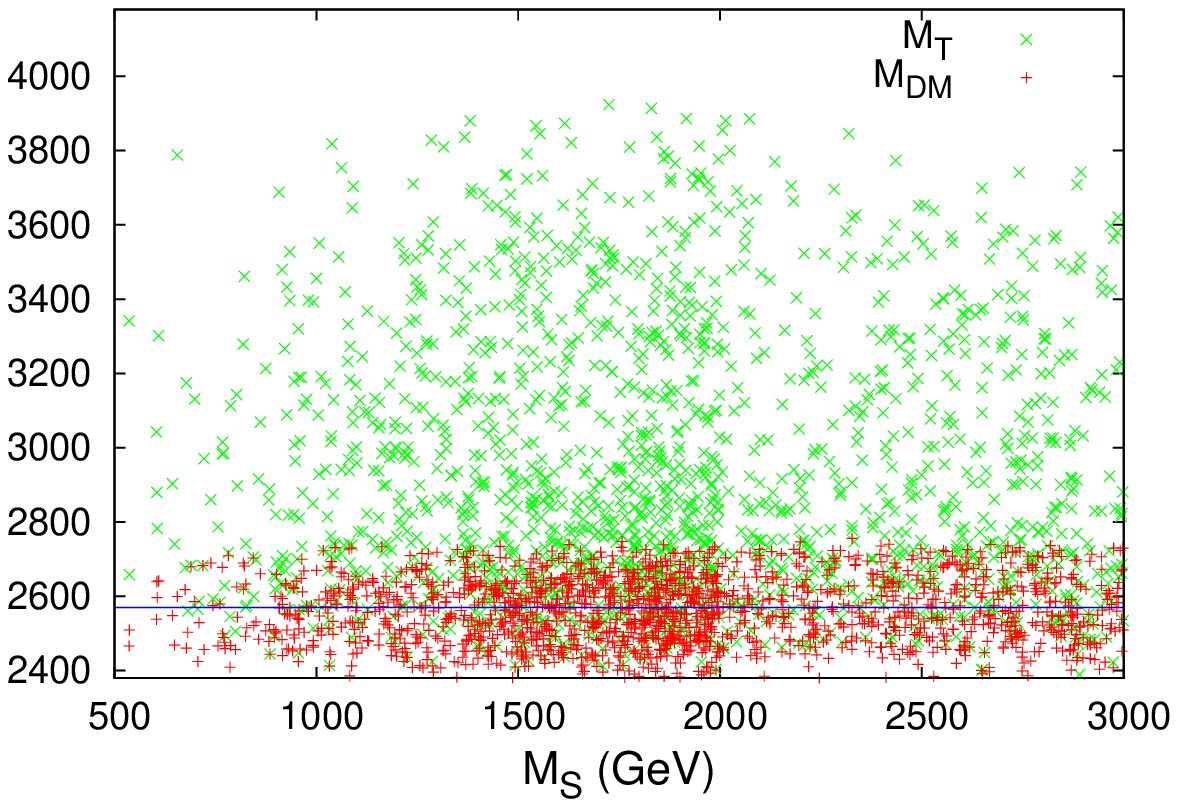}
\end{center}
\caption{Left: The DM relic density versus the freeze-out temperature,
$\Omega_{DM}h^{2}$ versus $x_{f}$, where $x_{f}=M_{DM}/T_{f}$. Right: the
allowed mass values where all the previous requirements are achieved. The blue
line represents the DM mass in the absence of the charged leptons annihilation
channel, i.e., $g_{i\alpha}\rightarrow0.$}%
\label{fig:Omega_vs_xf}%
\end{figure}

We find that the DM mass is of the same order of magnitude as previous results
in the literature~\cite{Ma:2008cu}. The $p$-wave annihilation channels into
charged leptons shift the DM mass from the value $M_{DM}\approx2.55~TeV$ (blue
line in the right panel of Figure \ref{fig:Omega_vs_xf}-right) into the range
$M_{DM}=2.35\sim2.75$~TeV, though the difference is not particularly large. If
the triplet fermion only contributes a fraction of the total DM abundance, the
requisite mass range increases accordingly. For example, for a triplet fermion
that only comprises 50\% of the observed relic abundance, the mass should lie
in the range $M_{DM}=3.3\sim3.8$~TeV.

In principle the cross sections are subject to a Sommerfeld enhancement due to
$SU(2)_{L}$ gauge boson exchange. However, at the freezeout temperature
$T_{f}\sim M_{DM}/25$, the electroweak symmetry is broken. When calculating
the Sommerfeld enhancement one should therefore consider massive mediators. It
is well known that the enhancement from a massive mediator turns off for
$M^{\prime}/(\alpha^{\prime}M_{DM})\gtrsim\mathcal{O}(1)$, where primes denote
the coupling and mass of the mediator (see e.g. Ref. \cite{Slatyer:2009vg}).
With $\alpha_{2}\approx1/30$, we have $M_{W}/(\alpha_{2}M_{DM})\approx1$, and
the enhancement gives a modest correction (less than $\mathcal{O}(1)$) that
can be neglected, at least in an initial treatment. Previous works found an
increase in $M_{DM}$ of roughly $500$~GeV, due to the
enhancement~\cite{Hisano:2006nn,Cirelli:2007xd}. We checked numerically that
with $M_{DM}\sim3$~TeV all other constraints can be satisfied, so a small
increase in $M_{DM}$ will not spoil our conclusions. A consistent treatment
requires both $s$-wave and $p$-wave enhancements~\cite{Cassel:2009wt} for a
massive mediator; such an analysis is beyond the scope of this paper.

\subsection{Direct Detection}

\begin{figure}[t]
\centering
\begin{tabular}
[c]{ccccc}%
\epsfig{file=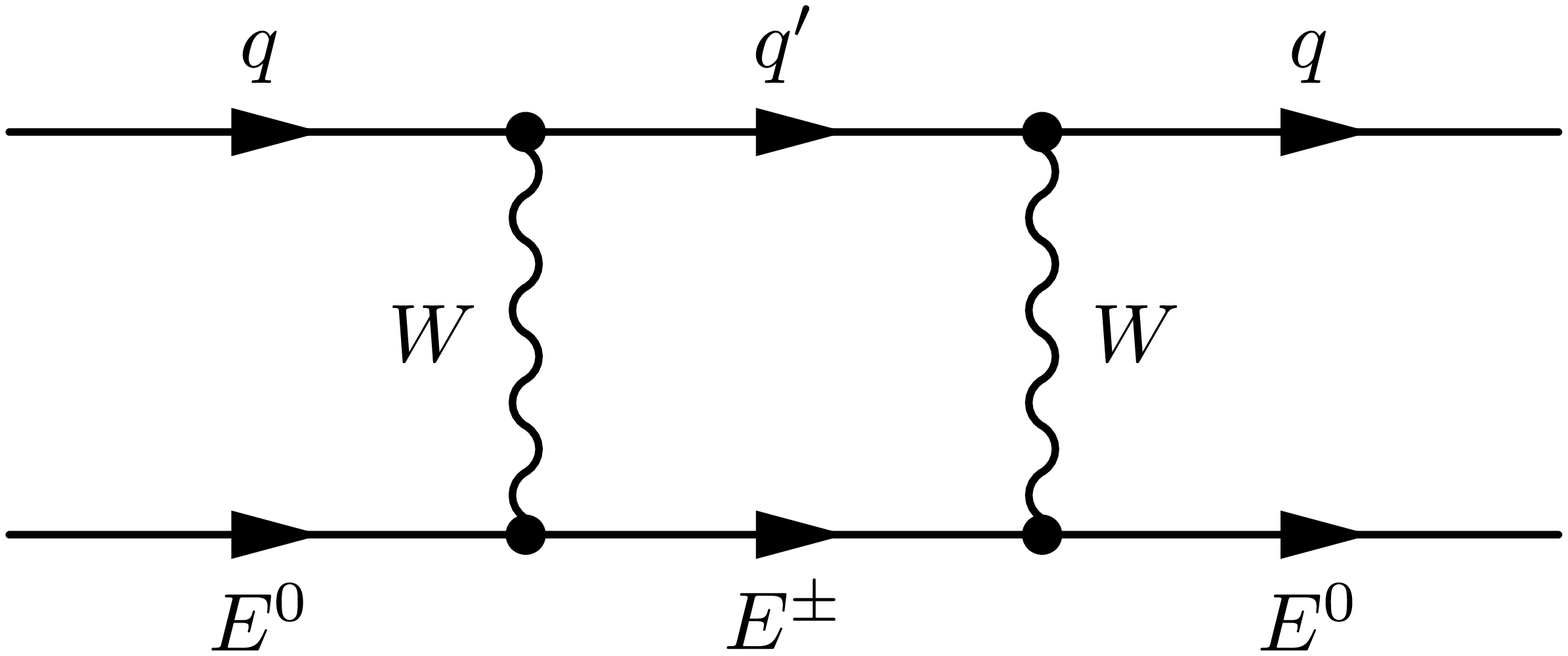,width=0.3\linewidth,clip=} & &
\epsfig{file=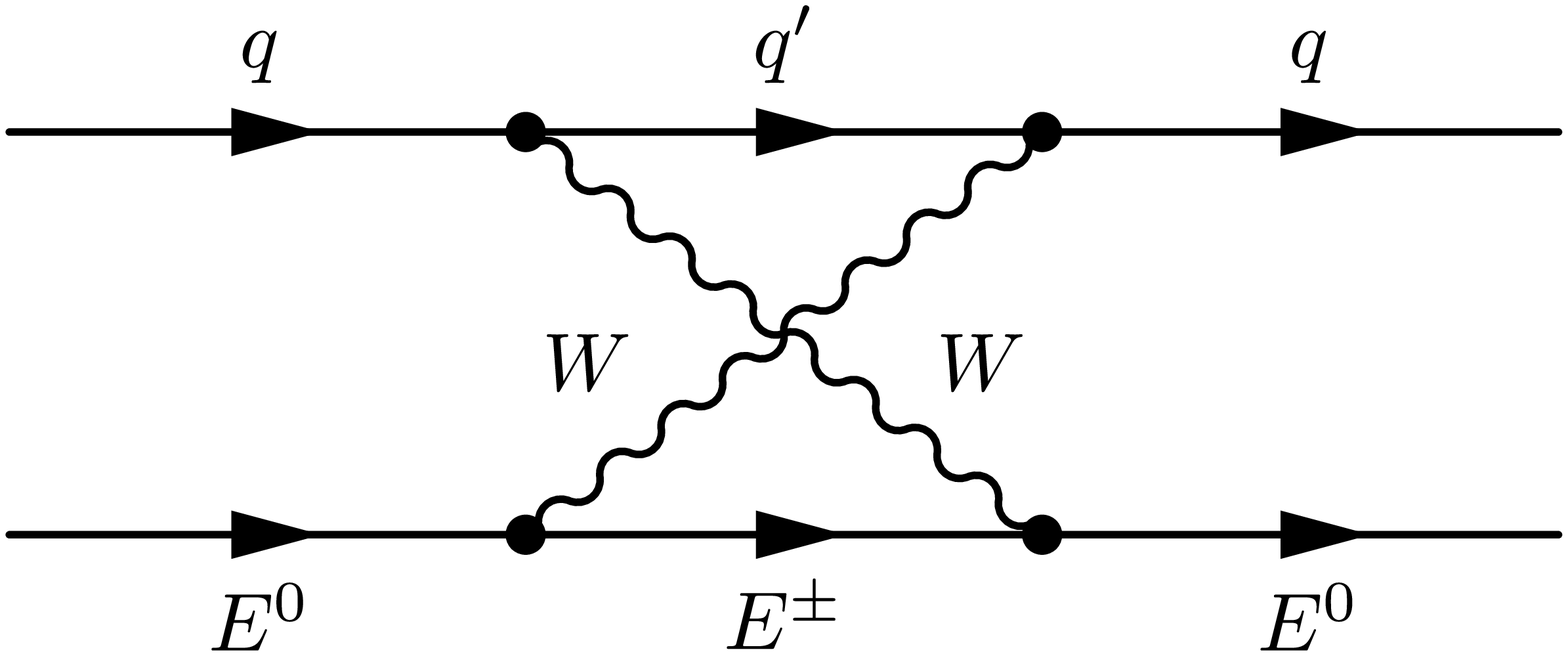,width=0.3\linewidth,clip=} & &
\epsfig{file=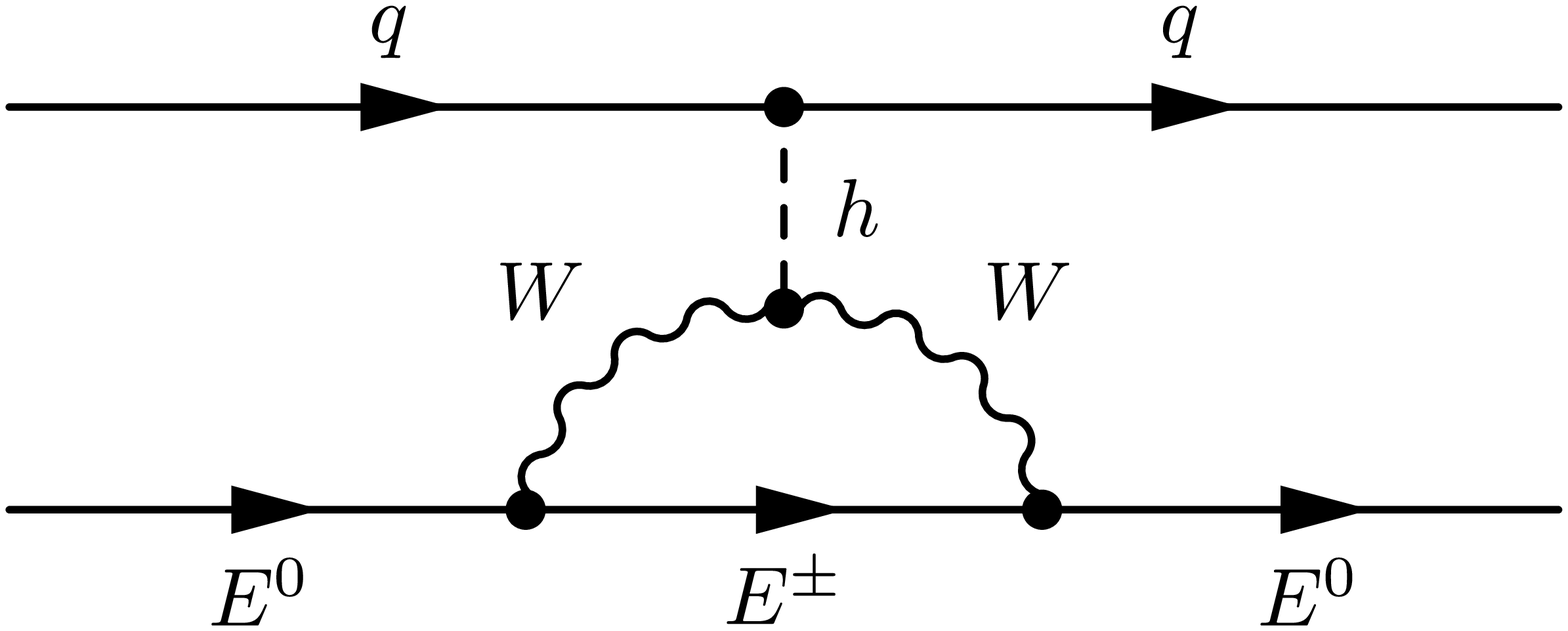,width=0.3\linewidth,clip=}\\
(a) & & (b) & & (c)\\
& & & &
\end{tabular}
\caption{Feynman diagrams relevant for direct-detection experiments.}%
\label{fig:DD_figure}%
\end{figure}

The triplet-fermion DM has vanishing isospin and consequently does not couple
to SM quarks at tree-level. Also, because the DM is a Majorana fermion, there
are no radiatively-induced magnetic dipole interactions with SM gauge bosons.
However, $W$ boson exchange generates three one-loop diagrams relevant for
direct-detection experiments; see Figure~\ref{fig:DD_figure}. The resultant
scattering has both spin-dependent and spin-independent contributions, though
the former are suppressed by the DM mass. The dominant interaction type is
therefore spin-independent scattering with cross section
\begin{equation}
\sigma_{\mathrm{SI}}(E^{0}N\rightarrow E^{0}N)\simeq\frac{\pi\alpha_{2}%
^{4}M_{A}^{4}f^{2}}{M_{W}^{2}}\left[ \frac{1}{M_{W}^{2}}+\frac{1}{M_{h}^{2}%
}\right] ^{2}.
\end{equation}
Here the DM scatters off a target nucleus $A$ with mass $M_{A}$, $\alpha_{2} $
is the $SU(2)_{L}$ fine-structure constant and we use the standard
matrix-element parametrization for the nucleon:
\begin{equation}
\langle N|\sum_{q}m_{q}\bar{q}q|N\rangle=fm_{N},
\end{equation}
with $m_{N}$ being the nucleon mass. We take $f\approx1/3$, though this is
subject to standard QCD uncertainties. This gives a cross section of roughly
$\sigma_{\mathrm{SI}}\simeq10^{-47}\mathrm{cm}^{2}$ per nucleon, which is
beyond the current sensitivity of experiments like LUX~\cite{Akerib:2013tjd},
but could be within reach of future experiments~\cite{superCDMS}. Future
prospects for probing this DM candidate are therefore promising.

\subsection{Indirect Detection}

Dark Matter candidates with non trivial electroweak quantum numbers can
generate observable signals at indirect-detection experiments. Though subject
to greater uncertainties than direct-detection experiments, indirect searches
can give useful constraints. For triplet-fermion DM, perhaps the strongest
indirect constraints come from gamma-ray searches from the Galactic center.
Gamma rays are produced via annihilations like $2E^{0}\rightarrow2\gamma$ and
$2E^{0}\rightarrow\gamma Z$, which occur at the one-loop level. The cross
sections are dominated by box diagrams with $W$ bosons and are largely
insensitive to the DM mass. The strongest constraints come from photon-line
searches of the Galactic center by Fermi-Lat~\cite{Ackermann:2013uma} and
HESS~\cite{Abramowski:2013ax}. The total rate for gamma-ray production depends
on the DM profile at the galactic center so the severity of the constraints
depends on the assumptions about the halo structure. Recent analyses of wino
DM in supersymmetric models~\cite{Cohen:2013ama,Fan:2013faa,Hryczuk:2014hpa}
find that thermal wino DM is consistent with the data for DM halos with a
significant core, while for the more cuspy Einasto and Navarro-Frenk-White
(NFW) distributions the parameter space with $10^{2}~\mathrm{GeV}\lesssim
M_{DM}\lesssim3$~TeV is excluded. For a cored profile like the Burkert 10~kpc
profile, a smaller region of parameter space with $2.25~\mathrm{TeV}\lesssim
M_{DM}\lesssim2.45$~TeV is excluded but wino DM remains otherwise
viable~\cite{Cohen:2013ama,Hryczuk:2014hpa}. Higher-order effects are expected
to weaken the bounds by a $\mathcal{O}(1)$ factor but this should not modify
the conclusion that thermal wino DM is excluded for the NFW
profile~\cite{Cohen:2013ama}. These constraints hold to good approximation for
the triplet-fermion DM in the present model. Together, the analysis suggests
that thermal DM requires a cored profile and a DM mass in the range
$M_{DM}\gtrsim2.45$~TeV. In the case where the triplet fermion does not supply
the full DM abundance these constraints are relaxed.

\section{Conclusion\label{sec:conc}}

We presented a new model in which the origin of neutrino mass is connected to
the existence of DM. The model is related to the KNT proposal but has a number
of distinguishing features. The DM should be somewhat heavy, $M_{DM}\sim$~TeV,
and can be probed in next-generation direct-detection experiments. We showed
that viable neutrino masses appear at the three-loop level and that
experimental constraints can be satisfied. Flavor changing effects like
$\mu\rightarrow e+\gamma$ could manifest at next-generation experiments, and
the model predicts a singly-charged scalar that can be within reach of
collider experiments. Discovery of the singlet scalar, in conjunction with the
observation of direct-detection signals from the triplet-fermion DM, would
provide strong evidence for this model. In a future work we shall study the
impact of the exotics on: the Higgs decay channels $h\rightarrow\gamma
\gamma,\gamma Z$, the possible enhancement of the triple-Higgs coupling, the
electroweak phase transition strength, and the collider phenomenology
\cite{next}, to determine if further signals are possible.

\section*{Acknowledgments\label{sec:ackn}}

A.A. thanks the ICTP for the hospitality where part of this work has
been carried out. The work of A.A. is supported by the Algerian
Ministry of Higher Education and Scientific Research under the
CNEPRU Project No. D01720130042. K.M. is supported by the Australian
Research Council. C.S.C. is supported by National Center for
Theoretical Sciences, Taiwan.

\appendix

\section{Triplet Fermion Couplings\label{app:chiral}}

In terms of this Majorana fermion $E^{0}=E_{L}^{0}-(E_{L}^{0})^{c}$, the
charged-current interactions take the form
\begin{equation}
\mathcal{L}_{W,E}=g\left( \overline{E^{+}}\gamma^{\mu}\,E^{0}\,W_{\mu}%
^{+}+\overline{E^{0}}\,\gamma^{\mu}E^{+}\,W_{\mu}^{-}\right) ,
\end{equation}
where $E^{+}=E_{L}^{+}+E_{R}^{+}$, while the coupling to the triplet scalar
is
\begin{align}
g_{i\alpha}\,(\overline{E_{i}})^{ab}\,T_{ba}\,e_{\alpha R} &
=g_{i\alpha
}\,\left\{ \overline{E_{iL}^{+}}\,T^{++}+\overline{E_{iL}^{0}}\,T^{+}%
+\overline{(E_{iR}^{+})^{c}}\,T^{0}\right\} \,e_{\alpha R}\nonumber\\
& =g_{i\alpha}\,\left\{ \overline{E_{i}^{+}}P_{R}\,e_{\alpha}\,T^{++}%
+\overline{E_{i}^{0}}\,P_{R}\,e_{\alpha}\,T^{+}+\overline{e_{\alpha R}^{v}%
}\,P_{R}\,E_{i}^{+}\,T^{0}\right\} \nonumber\\
& =g_{i\alpha}\,\left\{ \overline{E_{i}^{+}}P_{R}\,e_{\alpha}\,T^{++}%
-\overline{e_{\alpha}^{c}}\,P_{R}\,E^{0}\,T^{+}+\overline{e_{\alpha R}^{v}%
}\,P_{R}\,E_{i}^{+}\,T^{0}\right\} .
\end{align}
Note the extra negative sign in last form, which plays a role in the loop-mass
calculation. This sign difference results from the negative sign in the
Majorana mass terms in Eq.~\eqref{eq:mass_lagrangeE}.

\section{Radiative Neutrino Mass\label{app:loop_function}}

The Majorana neutrino masses are calculated to be
\begin{equation}
(\mathcal{M}_{\nu})_{\alpha\beta}=\frac{3\lambda_{S}}{(4\pi^{2})^{3}}%
\frac{m_{\gamma}m_{\delta}}{M_{T}}\,f_{\alpha\gamma}\,f_{\beta\delta
}\,g_{\gamma i}^{\ast}\,g_{\delta i}^{\ast}\times F\left( \frac{M_{i}^{2}%
}{M_{T}^{2}},\frac{M_{S}^{2}}{M_{T}^{2}}\right) ,
\end{equation}
where
\begin{equation}
F(\alpha,\beta)=\frac{\sqrt{\alpha}}{8\beta^{2}}\int_{0}^{\infty}dr\frac
{r}{r+\alpha}\left( \int_{0}^{1}dx\ln\frac{x(1-x)r+(1-x)\beta+x}%
{x(1-x)r+x}\right) ^{2}.
\end{equation}
In obtaining this form, the lepton masses that would otherwise
appear in the function $F$ have been neglected. The expression for
$F(\alpha,\beta)$ is the same as that found in the KNT
model~\cite{Ahriche:2013zwa}.

\end{document}